# The new $K_S \to \pi e \nu$ branching fraction measurement at KLOE

A. Passeri [a], on behalf of the KLOE and KLOE-2 Collaborations

[a] *INFN sezione Roma Tre, via della Vasca Navale 84, 00146 Roma, Italy*

*E-mail:* `antonio.passeri@roma3.infn.it`,

**Abstract**

A new measurement of the branching fraction for the decay $K_S \to \pi e \nu$ is presented, based on a sample of 300 million $K_S$ mesons recorded by the KLOE experiment at the DAΦNE $e^+e^-$ collider. A two-step signal selection strategy is used, exploiting first kinematic variables and then time-of-flight measurements. Data control samples of $K_L \to \pi e \nu$ decays are used to evaluate signal selection efficiencies. Normalizing the selected sample to the number of $K_S \to \pi^+\pi^-$ decay events the result for the branching fraction is $B(K_S \to \pi e \nu) = (7.211 \pm 0.046_{stat} \pm 0.052_{syst}) \times 10^{-4}$. The combination with our previous measurement gives $B(K_S \to \pi e \nu) = (7.153 \pm 0.037_{stat} \pm 0.043_{syst}) \times 10^{-4} = (7.153 \pm 0.057) \times 10^{-4}$. From this value we derive $f_+(0)|V_{us}| = 0.2170 \pm 0.0009$.

## 1. Introduction

The branching fraction for semileptonic decays of charged and neutral kaons together with the lifetime measurements are used to determine the |$V_{us}$| element of the Cabibbo–Kobayashi–Maskawa quark mixing matrix. Due to the lack of pure high-intensity $K_S$ meson beams compared with $K^\pm$ and $K_L$ mesons, the measurements of $K_S$ semileptonic decays from the KLOE [1, 2] and NA48 [3] experiments provide the least precise determination of |$V_{us}$|. We present here a new measurement of the $K_S \to \pi e \nu$ branching fraction performed by the KLOE experiment at the DAΦNE φ–factory [4] of the Frascati National Laboratory based on an integrated luminosity of 1.63 fb$^{-1}$.

The KLOE detector consists of a large Drift Chamber [5] and a Calorimeter [6], both immersed in a 0.5 T axial magnetic field. The cylindrical Drift Chamber (DC), with stereo wire geometry, provides ~ 150 μm single-hit spatial resolution in the bending plane and ~2 mm along the beam line, and allows charged particle tracks to be reconstructed with high momentum resolution ($\sigma_p/p = 0.4\%$). The Pb-scintillating fibers calorimeter (EMC) ensures a 98% solid angle coverage and allows clusters to be reconstructed with excellent time ($\sigma_t$ = 54 ps/$\sqrt{E(GeV)}$ ⊕ 100 ps) and good energy ($\sigma_E/E$ = 5.7%/$\sqrt{E(GeV)}$) resolutions. The level-1 trigger requires two energy deposits with E > 50 MeV in the barrel calorimeter and E > 150 MeV in the endcaps; the drift chamber trigger is based on the number and topology of hit drift cells. A higher-level logic rejects cosmic-ray events. An event sample equivalent to the data was simulated with the GEANFI simulation package [7]. Energy deposits in EMC and DC hits from beam background events triggered at random are overlaid onto the simulated events which are then processed with the same reconstruction algorithms as the data.

## 2. The branching fraction measurement

The branching fraction of the $K_S \to \pi e \nu$ decay is evaluated as

$$B(K_S \to \pi e \nu) = \frac{N_{\pi e \nu}}{\varepsilon_{\pi e \nu}} \times \frac{\varepsilon_{\pi\pi}}{N_{\pi\pi}} \times R_\varepsilon \times B(K_S \to \pi^+\pi^-) \qquad (1)$$

where $N_{\pi e \nu}$ and $N_{\pi\pi}$ are the numbers of selected $K_S \to \pi e \nu$ and $K_S \to \pi^+\pi^-$ events, $\varepsilon_{\pi e \nu}$ and $\varepsilon_{\pi\pi}$ are the respective selection efficiencies, and $R_\varepsilon = (\varepsilon_{\pi\pi}/\varepsilon_{\pi e \nu})_{com}$ is the ratio of common efficiencies for the trigger, on-line filter, event classification and preselection.

### 2.1 Signal selection and normalization sample

$K_S$ mesons are tagged by $K_L$ interactions in the calorimeter, $K_L$-crash in the following, with a clear signature of a delayed cluster not associated to tracks. To select $K_L$-crash and then tag $K_S$ mesons, the requirements are:
- one cluster not associated to tracks with energy $E_{clu}$ > 100 MeV, the centroid of the neutral cluster defining the $K_L$ direction with an angular resolution of ~1°;
- 15° < $\theta_{clu}$ < 165° for the polar angle of the neutral cluster, to suppress small-angle beam backgrounds;
- 0.17 < $\beta^*$ < 0.28 for the velocity in the $\phi$-meson system of the $K_L$ candidate;

$\beta^*$ is obtained from the velocity in the laboratory system, $\beta = r_{clu}/ct_{clu}$, with $t_{clu}$ being the cluster time and $r_{clu}$ the distance from the nominal interaction point. The $K_S$ momentum $\vec{p}_{K_S} = \vec{p}_\varphi - \vec{p}_{K_L}$ is determined with an accuracy of 2 MeV, assigning the neutral kaon mass.

$K_S \to \pi e \nu$ and $K_S \to \pi^+\pi^-$ preselection requires two tracks of opposite curvature forming a vertex inside the cylinder centered in the interaction point, with axis directed along the $z$ coordinate having 5 cm radius and 20 cm length. After preselection, the data sample contains about 300 million events, mainly $K_S \to \pi^+\pi^-$ decays, with 3% contamination of $\phi \to K^+K^-$ decays and only 0.08% of signal events, according to simulation.

Signal selection is performed in two steps based on uncorrelated information: 1) the event kinematics using only DC tracking variables, and 2) the time-of-flight measured with the EMC. Track-to-cluster association is required for both tracks, to allow time assignement. Associated clusters must have $E_{clu}$ > 20 MeV, 15° < $\theta_{clu}$ < 165° and must be within 30 cm of the track extrapolation to the calorimeter. A multivariate analysis is performed with a boosted decision tree (BDT) classifier built with the following variables with good discriminating power against background: the tracks' momenta, their opening angle, the angle between the total tracks momentum $\vec{p}_{sum}$ and the $K_L$-crash direction, the difference between $|\vec{p}_{sum}|$ and the $K_S$ momentum, the invariant mass of the two tracks in the pion mass hypothesis. The BDT classifier is trained with MC samples: 5,000 $K_S \to \pi e \nu$ events and 50,000 background events. Figure 1 shows the BDT output for data and simulated signal and background events. To suppress the large background contribution from $K_S \to \pi^+\pi^-$ and $\phi \to K^+K^-$ events, we then require BDT > 0.15.

Time-of-flight measurements are used to identify $e\pi$ pairs in the selected events. For each track associated to a cluster, the difference between the time-of-flight measured by the calorimeter and the flight time measured along the particle trajectory is computed: $\delta t = t_{clu} - L/c\beta$, where $t_{clu}$ is the time associated to the track, $L$ is the length of the track, and its velocity $\beta$ is function of the mass hypothesis for the particle that produced the observed track. To reduce the uncertainty from the determination of the event $T_0$ the difference of the $\delta t$ of the two tracks, $\delta t_{1,2}$ is used to determine the mass assignment.

The $\pi\pi$ hypothesis is tested first: $\delta t_{\pi\pi}$ distribution is shown in figure 2 and exhibits a fair agreement between data and simulation, with $K_S \to \pi e \nu$ and $K_S \to \pi \mu \nu$ distributions well separated and large part of the $K^+K^-$ background isolated in the tails of the distribution. The signal is hidden under a large $K_S \to \pi^+\pi^-$ background, therefore a cut 2.5 ns < $|\delta t_{\pi\pi}|$ < 10 ns is applied.

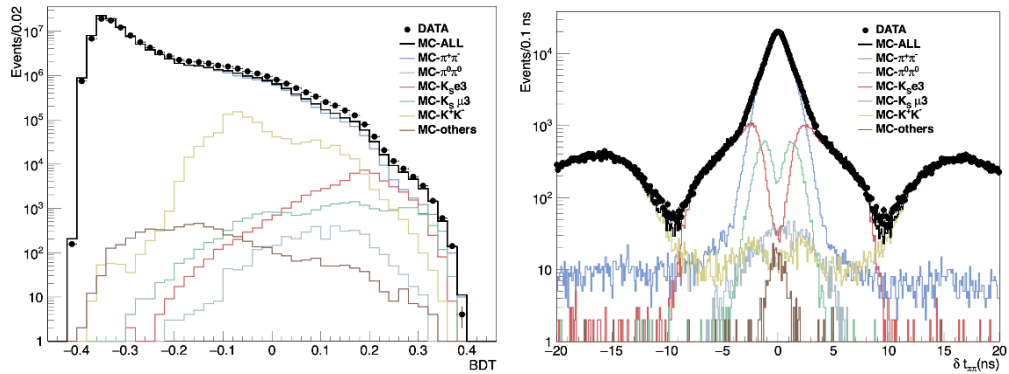

**Figure 1**. Distribution BDT classifier output (left) and of $\delta t_{\pi\pi}$ (right) for data, simulated signal and background events.

Then the $e\pi$ hypothesis is tested by assigning the pion and electron mass to either track, defining $\delta t_{e\pi} = \delta t_{1,e} - \delta t_{2,\pi}$ and $\delta t_{\pi e} = \delta t_{1,\pi} - \delta t_{2,e}$ where the label as track-1 and track-2 is chosen at random. Signal events have $\delta t \sim 0$ in one of the two hypotheses. If $\delta t_{\pi e} < \delta t_{e\pi}$ track-1 is assigned to the pion and track-2 to the electron, otherwise the other solution is taken; the lowest of the two time differences, $\delta t_e$, is then requited to be $|\delta t_e| < 1$ ns. The number of selected events is 57577, which according to simulation includes 94.22% of signal events, 3.83% of $K_S \to \pi^+\pi^-$ decays, 1.59% of $\phi \to K^+K^-$ events and 0.36% of other processes.

The mass of the charged secondary identified as the electron is evaluated as
$$m_e^2 = (E_{K_S} - E_\pi - p_{miss})^2 - p_e^2$$
with $p_{miss}^2 = (\vec{p}_{K_S} - \vec{p}_\pi - \vec{p}_e)^2$, $E_{K_S}$ and $\vec{p}_{K_S}$ being the energy and momentum reconstructed using the tagging $K_L$, and $\vec{p}_\pi, \vec{p}_e$ the momenta of the pion and electron tracks, respectively.

A fit to the $m_e^2$ distribution with the MC shapes of three components, $K_S \to \pi e \nu$, $K_S \to \pi^+\pi^-$ and the sum of all other backgrounds, allows the number of signal events to be extracted. The fit is performed in 100 bins in the range [-30000,+30000] MeV$^2$. Figure 3 shows the $m_e^2$ distribution for data and simulated events before the fit, and the comparison of the fit output with the data. The fitted number of signal events is:

$$N_{\pi e \nu} = 49647 \pm 316 \qquad \text{with } \chi^2 / \text{ndf} = 76/96.$$

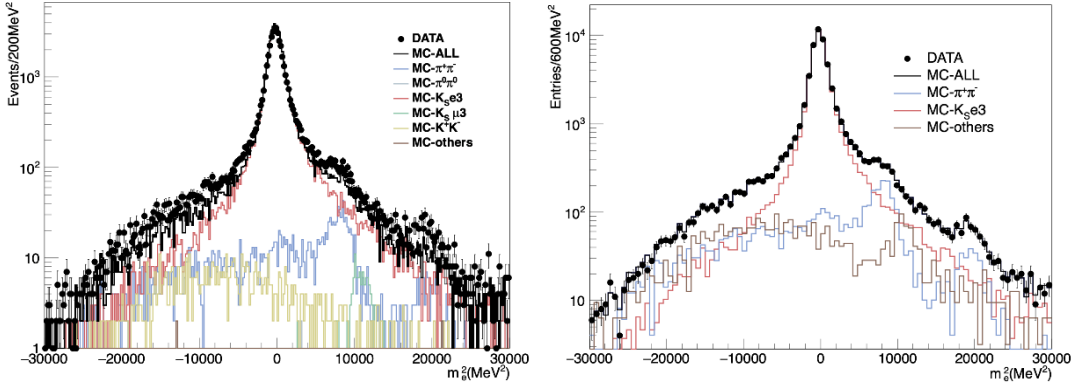

**Figure 3**. The $m_e^2$ distribution for data, MC signal and background before the fit (left) and comparison of data with the result of the fit (right).

The $K_S \rightarrow \pi^+\pi^-$ normalization sample is selected requiring $K_L$-crash, two opposite curvature tracks forming a vertex (with same geometrical selection as the signal vertexes) and $140 < p < 280$ MeV for both tracks. A total of $N_{\pi\pi} = (282.314 \pm 0.017) \times 10^6$ events are selected with an efficiency of 97.4% and a purity of 99.9% as determined using simulation.

### 2.2 Determination of efficiencies

The signal efficiency for a given selection is determined with a $K_L \rightarrow \pi e \nu$ control sample (CS) and evaluated as:

$$\varepsilon_{\pi e \nu} = \varepsilon_{CS} \times \frac{\varepsilon_{\pi e \nu}^{MC}}{\varepsilon_{CS}^{MC}} \qquad (2)$$

where $\varepsilon_{CS}$ is the efficiency of the control sample, and $\varepsilon_{\pi e \nu}^{MC}$, $\varepsilon_{CS}^{MC}$ are the efficiencies obtained from simulation for the signal and the control sample, respectively. Extensively studied with the KLOE detector [8], $K_L \rightarrow \pi e \nu$ decays are kinematically identical to the signal, the only difference being the much longer decay path. Tagging is done with $K_S \rightarrow \pi^+\pi^-$ decays selected requiring two opposite curvature tracks forming a vertex and having invariant mass (in the hypothesis they are both pions) within a ±15 MeV interval from the $K^0$ mass. The radial distance of the $K_L$ vertex is required to be smaller than 5 cm, to match the signal selection, but greater than 1 cm to minimize the ambiguity in identifying $K_L$ and $K_S$ vertices. $K_L$ decays in three pions are easily removed from the control sample.

Two high purity (>95%) sub-samples, each containing $O(10^6)$ events, have been then selected: the first applying a cut on the TOF variables to evaluate the efficiency of the selection based on the kinematic variables and the BDT classifier, the second applying a cut on kinematic variables to evaluate track-to-cluster association (TCA) and TOF selection efficiencies. Good data-MC agreement is observed for the distribution of the variables to be studied in each of the two control samples. Applying to both samples the same selections as for the signal, the efficiencies evaluated with Eq. (2) are reported in table 1.

For the $K_S \to \pi^+\pi^-$ normalization sample, the selection efficiency is evaluated exploiting the preselected data and is $\varepsilon_{\pi\pi} = (96.657 \pm 0.002)\%$, with a systematic uncertainty of 0.012% given by the difference between two different efficiency evaluation methods.

The ratio $R_\varepsilon$ in Eq. (1) includes several effects depending on the event global properties: trigger, on-line filter, event classification, $T_0$ determination, $K_L$-crash and $K_S$ identification, which are evaluated with simulation and give $R_\varepsilon = 1.1882 \pm 0.0012$ (statistical error only).

| Selection | Efficiency |
|---|---|
| Preselection (from MC) | 0.9961 ± 0.0002 |
| Kin.variables selection | 0.9720 ± 0.0007 |
| the BDT selection | 0.6534 ± 0.0013 |
| TCA selection | 0.4639 ± 0.0009 |
| TOF selection | 0.6605 ± 0.0012 |
| Total | 0.1938 ± 0.0006 |

**Table 1.** Signal selection efficiencies with statistical uncertainties. Correlations are accounted for in evaluating the total efficiency uncertainty.

### 3. Systematic uncertainties

The signal count is affected by three main systematic uncertainties: BDT selection, TOF selection, and the $m_e^2$ fit. The analysis is repeated varying the BDT cut in the range 0.135-0.17, observing good stability of the extracted number of signal events. Half width of its spread is taken as systematic uncertainty.

TCA efficiency calculation is repeated by weighting the events of the control sample by the number of track-associated clusters. The difference is less than 0.1% and is taken as relative systematic uncertainty. The $\delta t_e$ resolution has been checked to be the same in the signal and in the control sample. The lower $|\delta t_{\pi\pi}|$ cut has been varied in the range 2.0-3.0 ns while the $|\delta t_e|$ cut was varied in the range 0.8-1.2 ns. In both cases the half-width of the band was taken as systematic uncertainty, respectively ±0.28% and ±0.12%.

The fit to the $m_e^2$ distribution has been repeated varying the range and the bin size, and also using two separate components for $K_S \to \pi\mu\nu$ and $\phi \to K^+K^-$. Half of the difference between maximum and minimum result of the different fits, 0.15%, is taken as relative systematic uncertainty. All systematic uncertainties are listed in Table 2.

The systematic effects related to $R_\varepsilon$ have been studied in previous analyses and are evaluated by a comparison of data with simulation. The difference from one of the Data/MC ratio is taken as systematic uncertainty. The combined statistical and systematic uncertainty on $R_\varepsilon$ is 0.0059.

| Selection | $\delta\varepsilon_{\pi e\nu}^{syst}$ [$10^{-4}$] | $\delta\varepsilon_{\pi^+\pi^-}^{syst}$ [$10^{-4}$] |
|---|---|---|
| BDT selection | 5.3 | |
| TCA & TOF selection | 6.0 | |
| Fit parameters | 3.0 | |
| $K_S \to \pi^+\pi^-$ efficiency | | 8.8 |
| Total | 8.5 | 8.8 |

**Table 2.** Systematic uncertainties of efficiencies.

## 4. The Result

Using Eq.(1) with all the measured number of events and efficiencies, and $B(K_S\rightarrow\pi^+\pi^-) = 0.69196 \pm 0.00051$ measured by KLOE [9] we derive the branching fraction:

$$B(K_S \rightarrow \pi e\nu) = (7.211 \pm 0.046_{stat} \pm 0.052_{syst}) \times 10^{-4} = (7.211 \pm 0.069) \times 10^{-4}.$$

The combination with the previous result from KLOE [6], based on an independent data sample corresponding to 0.41 fb$^{-1}$ of integrated luminosity, accounting for correlations between the two measurements, gives

$$B(K_S \rightarrow \pi e\nu) = (7.153 \pm 0.037_{stat} \pm 0.043_{syst}) \times 10^{-4} = (7.153 \pm 0.057) \times 10^{-4}.$$

The value of $|V_{us}|$ is related to the $K_S$ semileptonic branching fraction by the equation

$$B(K_s \rightarrow \pi l\nu) = \frac{G^2(f_+(0)|V_{us}|)^2}{192\pi^3} \tau_S m_K^5 I_K^l S_{EW}(1 + \delta_{EM}^{Kl})$$

where $I_K^l$ is the phase-space integral, which depends on measured semileptonic form factors, $S_{EW}$ is the short-distance electro-weak correction, $\delta_{EM}^{Kl}$ is the mode-dependent long-distance radiative correction, and $f_+(0)$ is the form factor at zero momentum transfer for the $l\nu$ system. Using the values of $S_{EW}$ from ref.[10], $I_K^l$ and $\delta_{EM}^{Kl}$ from ref.[11] and the world average values for the $K_S$ mass and lifetime [12] we derive

$$f_+(0)|V_{us}| = 0.2170 \pm 0.0009$$

which has an accuracy better than that which can be derived from the KLOE measurements of $K^\pm$ semileptonic decays branching fractions [13].

Finally, the sum of the main $K_S$ branching fractions, all measured by KLOE ([9],[2] and the present result), yields:

$$B_{\pi^+\pi^-} + B_{\pi^0\pi^0} + B_{\pi e\nu} + B_{\pi\mu\nu} > 0.9983 \text{ @95\% CL}$$